\documentclass[12pt,preprint]{aastex}

\begin{document}

\title{The first photometric analysis of the overcontact binary MQ UMa with an additional component}

\author{X. Zhou\altaffilmark{1,2,3}, S.-B. Qian\altaffilmark{1,2,3},W.-P. Liao\altaffilmark{1,2},E.-G. Zhao\altaffilmark{1,2,3},J.-J. Wang\altaffilmark{1,2,3},L.-Q. Jiang\altaffilmark{1,2,3}}

\singlespace

\altaffiltext{1}{Yunnan Observatories, Chinese Academy of Sciences (CAS), P. O. Box 110, 650216 Kunming, China; zhouxiaophy@ynao.ac.cn}
\altaffiltext{2}{Key Laboratory of the Structure and Evolution of Celestial Objects, Chinese Academy of Sciences, P. O. Box 110, 650216 Kunming, China}
\altaffiltext{3}{University of the Chinese Academy of Science, Yuquan Road 19\#, Sijingshang Block, 100049 Beijing, China}

\begin{abstract}
The first $V$ $R_c$ $I_c$ bands light curves of MQ UMa are presented and analyzed using the Wilson-Devinney (W-D) program. It is discovered that MQ UMa is an A-subtype contact binary with a high fill-out (f=$82\,\%$) and a low mass ratio ($q=0.195$), which indicates that it is in the late evolutionary stage of late-type tidal-locked binary stars. The mass of the primary and secondary stars are estimated and the evolutionary status of the two components are placed in the H-R diagram. The W-D solutions also indicates that MQ UMa may be a triple system with an F5V type additional component. A sample of sixteen high fill-out, low mass ratio overcontact binaries are collected and their possible evolution scenarios are discussed. Using the five times of light minimum recently observed together with those collected from literatures, the authors find that the observed-calculate ($O$-$C$) curve exhibits a cyclic period variation. The cyclic period change also reveals the presence of a tertiary component, which may play an important role in the formation and evolution of this binary system by drawing angular momentum from the central system.

\end{abstract}


\section{Introduction}
W UMa type binaries are cool short-period (usually less than 1 day) binary systems with both components filling their critical Roche lobes and sharing a common convective envelope during their main sequence evolutionary stage. The formation and evolution of W UMa type binary systems are still unsolved problems in astrophysics. The most popular evolutionary scenario is that they are formed from initially detached systems via angular momentum loss (AML) by means of magnetic stellar wind \citep{1982A&A...109...17V,1989AJ.....97..431E}. Model calculations suggest that these binary stars will ultimately coalesce into single stars which may be progenitors of the poorly understood blue stragglers and FK Com stars \citep{2006AcA....56..199S,2011AcA....61..139S}. It has been widely accepted that the eruption of V1309 Sco was the result of a cool short-period binary merging. In this paper, we focus on the high fill-out, low mass ratio overcontact binaries which are at the late evolutionary stages of the contact configuration. Photometric analysis and orbital period studies of them will provide important information for the evolution and coalescence scenario of these binary systems.

MQ UMa, also named GSC 3015 0374, is a typical W UMa type contact binary. It was first discovered by CCD observations from large-scale automatic sky surveys in 1999 \citep{2005PZ.....25....2K}. Since then, several times of light minimum have been published, and it is included in the Tycho-2 Catalogue and 2MASS All Sky Catalogue. The Tycho-2 Catalogue gives the magnitude of MQ UMa, which are 11.77 (0.09) mag in $B$ band and 11.22 (0.09) mag in $V$ band. The 2MASS All Sky Catalogue gives the magnitude of MQ UMa in $J$, $H$ and $K$ band filters, which are 10.616 (0.026) mag in $J$ band, 10.402 (0.028) mag in $H$ band and 10.387 (0.018) mag in $K$ band. However, there is neither light curve (LC) photometric solution nor spectroscopic information about this target.

\section{The CCD photometric light curves and times of light minimum}
The $V$ $R_c$ and $I_c$ bands CCD observations of MQ UMa were carried out in three nights on January 19, March 7 and  April 26, 2014 with an Andor DW436 1K CCD camera attached to the 85cm reflecting telescope at Xinglong Observation Base. The coordinates of the variable star, the comparison star and the check star were listed in Table \ref{Coordinates}. During the observation, the broad band, Johnson-Cousins $V$ $R_c$ $I_c$ filters were used. The integration time were 60s for $V$ band, 40s for $R_c$ band, and 30s for $I_c$ band, respectively. PHOT (measure magnitudes for a list of stars) of the aperture photometry package in the IRAF \footnote {The Image Reduction and Analysis Facility is hosted by the National Optical Astronomy Observatories in Tucson, Arizona at URL iraf.noao.edu.} was used to reduce the observed images. The average observational errors were 0.002 mag for $V$ band, 0.003 mag for $R_c$ band and 0.003 mag for $I_c$ band, respectively. The light curves of those observations were displayed in Fig. 1.

\begin{table}[!h]
\begin{center}
\caption{Coordinates of MQ UMa, the comparison, and the check stars.}\label{Coordinates}
\begin{small}
\begin{tabular}{cccc}\hline\hline
Targets          &   name            & $\alpha_{2000}$        &  $\delta_{2000}$   \\ \hline
Variable         &   MQ UMa          &$11^{h}21^{m}41^{s}.1$  & $+43^\circ36'52''.8$\\
The comparison   &   GSC 3015 0367  &$11^{h}21^{m}19^{s}.2$  & $+43^\circ38'09''.9$\\
The check        &   GSC 3015 0408	 &$11^{h}21^{m}19^{s}.2$  & $+43^\circ31'51''.4$\\
\hline\hline
\end{tabular}
\end{small}
\end{center}
\end{table}

\begin{figure}[!h]
\begin{center}
\includegraphics[width=13cm]{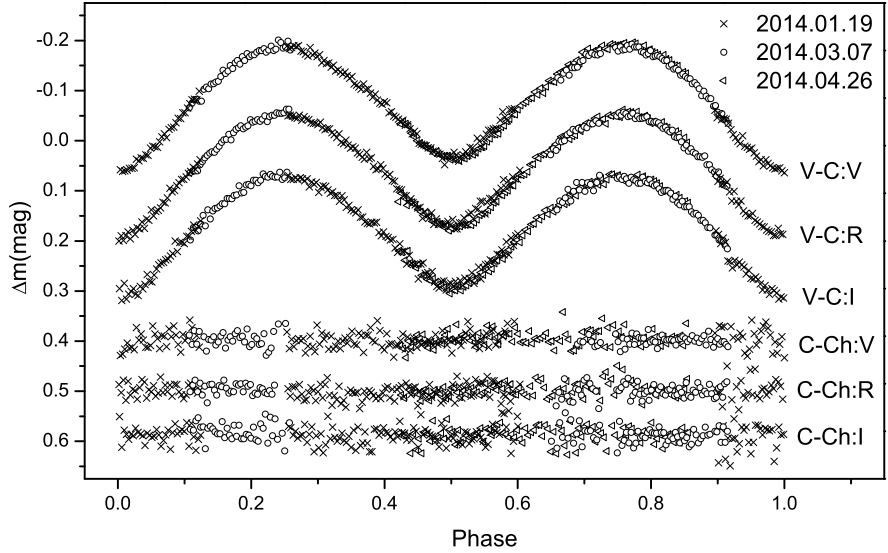}
\caption{CCD photometric light curves in $V$ $R_c$ and $I_c$ bands. The magnitude difference between the comparison and the check stars are presented. The standard deviations of the comparison-check observations are 0.015 mag for $V$ band, 0.015 mag for $R_c$ band and 0.017 mag for $I_c$ band. Crosses, open circles and triangles correspond to the data observed on January 19, March 7 and April 26, respectively.}
\end{center}
\end{figure}

We also got three times of light minimum on January 19 and  April 26, 2014 while doing light curve observations. After that, one time of light minimum was obtained on January 3, 2015 using the 60cm reflecting telescope in Yunnan Observatories (YNOs), and another TOM was obtained on January 12, 2015 using the 1m reflecting telescope in YNOs. Basing on the least-square parabolic fitting method, five new CCD times of light minimum were determined and listed in Table \ref{Newminimum}.

\begin{table}[!h]
\begin{center}
\caption{New CCD times of light minimum for MQ UMa.}\label{Newminimum}
\begin{tabular}{cccccc}\hline
    JD (Hel.)     &  Error (days)  & Min. &   Filter    & Method &Telescopes\\\hline
  2456677.1653   & $\pm0.0004$   &   I  &   $VR_cI_c$ &  CCD   &    85cm   \\
  2456677.4031   & $\pm0.0005$   &   II &   $VR_cI_c$ &  CCD   &    85cm   \\
  2456774.0451   & $\pm0.0003$   &   II &   $VR_cI_c$ &  CCD   &    85cm   \\
  2457026.3616   & $\pm0.0006$   &   II &   $VR_cI_c$ &  CCD   &    1m   \\
  2457035.4084   & $\pm0.0004$   &   II &   $I_c$     &  CCD   &    60cm   \\
\hline
\end{tabular}
\end{center}
\textbf
{\footnotesize Notes.} \footnotesize 60cm and 1m denote to the 60cm and 1m reflecting telescope in Yunnan Observatories, and 85cm denotes to the 85cm reflecting telescope in Xinglong Observation base.
\end{table}

\section{Orbital period change of MQ UMa}

The study of orbital period change is a very important part for contact binary stars. However, the period change investigation of MQ UMa has been neglected since it was discovered. During the present work, all available times of light minimum are collected and listed in Table \ref{Minimum}. Using the following linear ephemeris,
\begin{equation}
Min.I(HJD)=2451312.851+0^{d}.476066\times{E},\label{linear ephemeris}
\end{equation}
the $O - C$ (observational times of light minimum-calculational times of light minimum) values are calculated and listed in the fourth column of Table \ref{Minimum} and plotted in the upper panel of Fig 2. Minimum times with the same epoch have been averaged, and only the mean values are listed in Table \ref{Minimum}. The general $O-C$ trend of MQ UMa shown in the upper panel of Fig. 2 indicates a cyclic change in its orbital period. Based on the least-square method, a sinusoidal term is added to the linear ephemeris of Equation (2). The new ephemeris is
\begin{equation}
\begin{array}{lll}
Min. I=2451312.85732(\pm0.00029)+0.47606620(\pm0.00000003)\times{E}
         \\+0.005646(\pm0.000135)\sin[0.^{\circ}03438\times{E}+264.^{\circ}293(\pm1.^{\circ}356)]
\end{array}
\end{equation}
The sinusoidal term reveals a cyclic change with a period of 13.6 years and an amplitude of 0.0056 days. The residuals from Equation (2) are displayed in the lowest panel of Fig. 2.

\begin{table*}[!h]
\caption{$(O-C)$ values of light minima for MQ UMa.}\label{Minimum}
 \begin{center}
 \small
   \begin{tabular}{lclllcc}\hline\hline
JD (Hel.)      &  Min     &   Epoch      & $(O-C)$        &   Error   & Method    &  Reference       \\
(2400000+)     &          &              &                &           &           &\\\hline
51312.8513     & I        & 0            & 0.0003         &---        &  CCD      & 1 \\
54192.5885     & I        & 6049         & 0.0143         &0.0008     &  CCD      & 2 \\
54499.6476     & I        & 6694         & 0.0108         &0.0030     &  CCD      & 3 \\
54518.9277     & II       & 6734.5       & 0.0102         &0.0005     &  CCD      & 4 \\
54912.3940     & I        & 7561         & 0.0080         &0.0001     &  CCD      & 5 \\
54931.4375     & I        & 7601         & 0.0088         &0.0007     &  CCD      & 6 \\
55259.4448     & I        & 8290         & 0.0066         &0.0003     &  CCD      & 7 \\
55289.4355     & I        & 8353         & 0.0052         &0.0003     &  CCD      & 7 \\
55311.5753     & II       & 8399.5       & 0.0079         &0.0025     &  CCD      & 8 \\
55625.2975     & II       & 9058.5       & 0.0027         &0.0010     &  CCD      & 9 \\
55644.5823     & I        & 9099         & 0.0068         &0.0003     &  CCD      & 10 \\
55660.5289     & II       & 9132.5       & 0.0052         &0.0034     &  CCD      & 11 \\
55669.5757     & II       & 9151.5       & 0.0067         &0.0004     &  CCD      & 10 \\
55877.6101     & II       & 9588.5       & 0.0003         &0.0005     &  CCD      & 9 \\
55937.8357     & I        & 9715         & 0.0035         &0.0004     &  CCD      & 12 \\
56003.5322     & I        & 9853         & 0.0029         &0.0020     &  CCD      & 9 \\
56003.5325     & I        & 9853         & 0.0032         &0.0003     &  CCD      & 9 \\
56011.6263     & I        & 9870         & 0.0039         &0.0004     &  CCD      & 13 \\
56677.1653     & I        & 11268        & 0.0026         &0.0004     &  CCD      & 14\\
56677.4031     & II       & 11268.5      & 0.0023         &0.0004     &  CCD      & 14\\
56774.0451     & II       & 11471.5      & 0.0030         &0.0003     &  CCD      & 14\\
57026.3616     & II       & 12001.5      & 0.0042         &0.0006     &  CCD      & 14\\
57035.4084     & II       & 12020.5      & 0.0057         &0.0004     &  CCD      & 14\\\hline
\end{tabular}
\end{center}
\textbf
{\footnotesize Reference:} \footnotesize (1) \citet{2005PZ.....25....2K}; (2) \citet{2009IBVS.5874....1H}; (3) \citet{2008OEJV...94....1B}; (4) \citet{2009IBVS.5875....1N}; (5) \citet{2009OEJV..107....1B}; (6) \citet{2010IBVS.5918....1H}; (7) \citet{2011OEJV..137....1B}; (8) \citet{2011IBVS.5959....1H}; (9) \citet{2013OEJV..160....1H}; (10) \citet{2013IBVS.6070....1H};
(11) \citet{2012IBVS.6010....1H}; (12) \citet{2013IBVS.6050....1N}; (13) \citet{2013IBVS.6084....1H}; (14) present work;
\end{table*}

\begin{figure}[!h]
\begin{center}
\includegraphics[width=13cm]{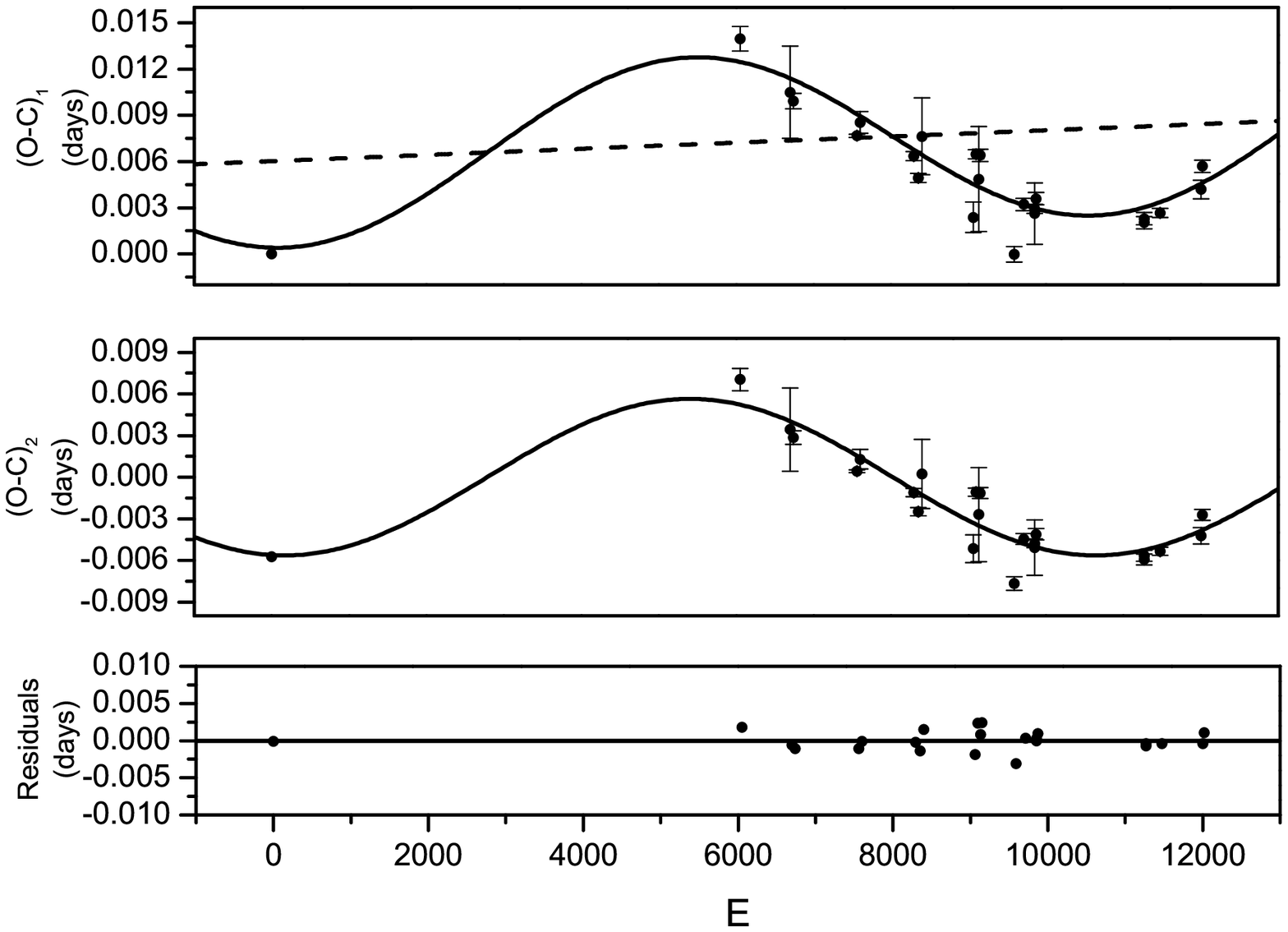}
\caption{The $(O-C)_1$ values of MQ UMa from the linear ephemeris of Equation (1) is presented in the upper panel. The solid line in the panel refers to the combination of a new linear ephemeris and a cyclic variation. The dash line represents the new linear ephemeris. In the middle part of Fig. 2, $(O-C)_2$ values calculated from the new linear ephemeris in Equation (2) are displayed. The solid line refers to a theoretical light travel time effect (LTTE) orbit of the tertiary component in the system. The residuals from the whole effect are displayed in the lowest panel.}
\end{center}
\end{figure}

It has to be mentioned that the data point at E = 0 seriously affect the $O - C$ fitting since there is no other time of light minimum between data point at E = 0 and E = 6049. Thus we check the first data carefully. Although the data point at E = 0 did not give its observational error, it was observed by CCD camera. We believe that it has a high time precision as other CCD data. So we add it to the $O - C$ fitting and give the results as above. The cyclic period change in Equation (2) may be caused by the light travel time effect (LTTE) of the third component. Actually, we can only ensure the existence of a tertiary component. The exact orbit period of the third body can not be determined  for the absence of data point between E = 0 and E = 6049. In the present work, we just estimate a typical period which gives a very nice fitting results in Fig. 2. In order to verify the periodic variations presented here, more determinations of light minima are required in the future.

\section{Photometric solutions of MQ UMa}

MQ UMa is a newly determined binary system, and neither light curve photometric solution nor spectroscopic observation has been published since it was discovered in 1999 \citep{2005PZ.....25....2K}. As shown in Fig. 1, the light curve variations in three colors are continuous and have very small magnitude differences between the depth of the primary and secondary minima. It means that MQ UMa ia a typical EW-type contact binary. In Fig.1, the light curve data have been shifted vertically which will make no difference to the results of W-D program as differential photometry method is used. The phases are calculated with the following linear ephemeris:
\begin{equation}
Min.I(HJD)=2456677.1653(4)+0^{d}.476058\times{E}.\label{linear ephemeris}
\end{equation}
To understand its geometrical structure and evolutionary state, the $V$ $R_c$ and $I_c$ light curves shown in Fig. 1 are analyzed using the W-D program of 2013 version \citep{Wilson1971,Wilson1979,Wilson1990,Van2007,Wilson2008,Wilson2010,Wilson2012}. The number of observational data points used in W-D program are 377 in $V$ band, 380 in $R_c$ band, 375 in $I_c$ band, respectively. According to the Tycho-2 Catalogue measurements, the color index of $B - V =0.55$, which corresponds to a spectral type of F9, but the 2MASS All Sky Catalogue gives the color index of $J - H = 0.214$, corresponding to a spectral type of F4 \citep{Cox2000}. Thus the spectral type of MQ UMa ranges from F9 to F4, which means the effective temperature of the primary star ranges from 6095K to 6670K \citep{Cox2000}. Meanwhile, we also use the following period - color relation derived by \citet{Deb2011}  to estimate the effective temperature of the primary star.
\begin{equation}
\begin{array}{lll}
$J - K$ = (0.11\pm0.01)\times{P^{-1.19\pm0.08}}
\end{array}
\end{equation}
The equation is derived from a total sample of 141 contact binaries, whose spectral type ranges from A2V to K5V, and has a period from 0.2211 days to 1.1318 days. The sample has covered nearly all kinds of W UMa type contact binaries which means it is appropriate to MQ UMa. The period of MQ UMa is 0.476058 days, so the color index of $J - K$ calculated by Equation (4) is 0.266 $(\pm0.042)$, which corresponds to spectral type of F5 to F7. Because of the advantage of $J - K$ color is not affected by interstellar extinction and uncertain reddening corrections as compared to $B - V$ color, the period-color relation using the infrared color $J - K$ will be much more accurate. According to the temperature estimated by the two merhods, the authors argue that MQ UMa is a late-F type W UMa contact binary system and adopt F7V as the spectral type of the primary star.

During the W-D processing, the effective temperature of star 1 is chosen as $T_1=6352$K according to the spectral type determined. Convective outer envelopes for both components are assumed. The bolometric albedo $A_1=A_2=0.5$ \citep{1969AcA....19..245R} and the values of the gravity-darkening coefficients $g_1=g_2=0.32$ \citep{1967ZA.....65...89L} are used. To account for the limb darkening in detail, logarithmic functions are used. The corresponding bolometric and passband-specific limb-darkening coefficients are chosen from \citet{1993AJ....106.2096V}'s table.
During the calculating, it is find that the solution converges at mode 3, and the adjustable parameters are: the mass ratio $q$ $(M_2/M_1)$; the orbital inclination $i$; the mean temperature of star 2 ($T_{2}$); the monochromatic luminosity of star 1 ($L_{1V}$, $L_{1R}$ and $L_{1I}$); the dimensionless potential of star 1 ($\Omega_{1}=\Omega_{2}$ in mode 3 for overcontact configuration); and the third light ($l_3$). Since there is no radial velocity curves of MQ UMa, a $q$-search method is used to determine the initial mass ratio at first. Solutions with mass ratio from 0.1 to 8 are investigated, and the relation between the resulting sum of weighted square deviations $\Sigma$ and $q$ is plotted in Fig. 3. The minimum values are found at $q$ = 0.2, which indicates that MQ UMa is an A-subtype contact binary. Then $q$ = 0.2 is set as the initial value and considered as an adjustable parameter. The final photometric solutions are listed in Table 4 and the theoretical light curves (with $l_3$) are displayed in Fig. 4. The contact configuration of MQ UMa is displayed in Fig. 5.

\begin{figure}[!h]
\begin{center}
\includegraphics[width=13cm]{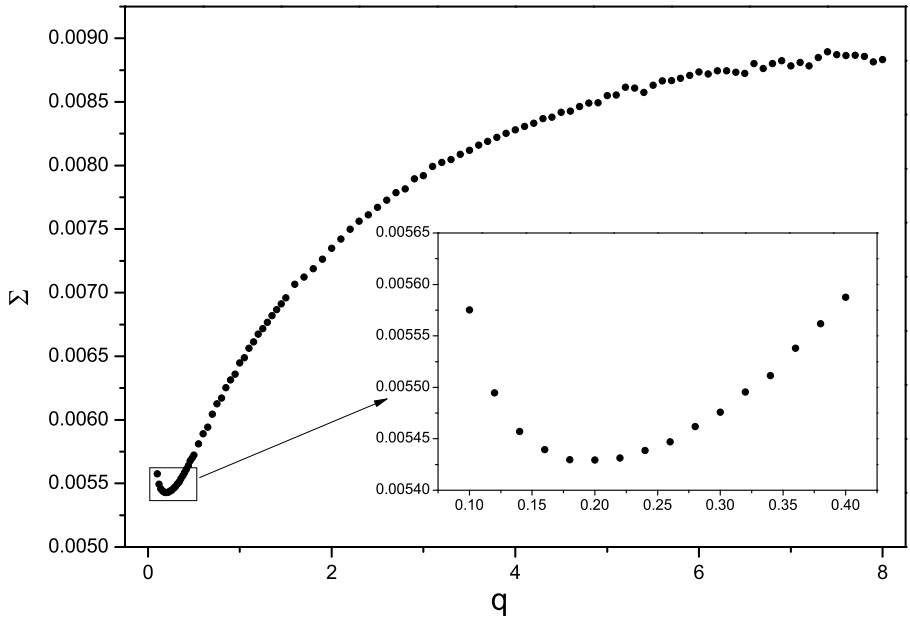}
\caption{Relation between $\Sigma$ and $q$ for MQ UMa. $\Sigma$ is the resulting sum of weighted square deviations. It is shown that the minimum is at q = 0.20.}
\end{center}
\end{figure}

\begin{table}[!h]
\caption{Photometric solutions of MQ UMa}\label{phsolutions}
\begin{center}
\small
\begin{tabular}{lllllllll}
\hline
Parameters                        & Values                        & Values       \\
                                  & without $l_{3}$               & with $l_{3}$   \\
\hline
$g_{1}$                           & 0.32(fixed)                   & 0.32(fixed)  \\
$g_{2}$                           & 0.32(fixed)                   & 0.32(fixed) \\
$A_{1}$                           & 0.50(fixed)                   & 0.50(fixed)   \\
$A_{2}$                           & 0.50(fixed)                   & 0.50(fixed)       \\
q ($M_2/M_1$ )                    & 0.211($\pm0.006$)             & 0.195($\pm0.008$)   \\
$T_{1}(K)   $                     & 6352(fixed)                   & 6352(fixed)         \\
$i(^{\circ})$                     & 60.70($\pm0.20$)              & 65.58($\pm0.69$)      \\
$\Omega_{in}$                     & 2.259129                      & 2.622506        \\
$\Omega_{out}$                    & 2.125046                      & 2.393652             \\
$\Omega_{1}=\Omega_{2}$           & 2.203736($\pm0.001206$)       & 2.117244($\pm0.021541$)           \\
$T_{2}(K)$                        & 6116($\pm12$)                 & 6224($\pm25$)        \\
$L_{1}/(L_{1}+L_{2}$) (V)         & 0.8211($\pm0.0009$)           & 0.8089($\pm0.0065$)   \\
$L_{1}/(L_{1}+L_{2}$) (R)         & 0.8172($\pm0.0009$)           & 0.8068($\pm0.0064$)   \\
$L_{1}/(L_{1}+L_{2}$) (I)         & 0.8138($\pm0.0009$)           & 0.8050($\pm0.0065$)   \\
$L_{1}/(L_{1}+L_{2}+L_{3}$) (V)   &                               & 0.5970($\pm0.0102$)\\
$L_{1}/(L_{1}+L_{2}+L_{3}$) (R)   &                               & 0.6057($\pm0.0099$)\\
$L_{1}/(L_{1}+L_{2}+L_{3}$) (I)   &                               & 0.6103($\pm0.0100$)\\
$L_{3}/(L_{1}+L_{2}+L_{3}$) (V)   &                               & 0.2620($\pm0.0068$)\\
$L_{3}/(L_{1}+L_{2}+L_{3}$) (R)   &                               & 0.2492($\pm0.0064$)\\
$L_{3}/(L_{1}+L_{2}+L_{3}$) (I)   &                               & 0.2419($\pm0.0064$)\\
$r_{1}(pole)$                     & 0.4963($\pm0.0023$)           & 0.5150($\pm0.0038$)   \\
$r_{1}(side)$                     & 0.5436($\pm0.0031$)           & 0.5710($\pm0.0055$)      \\
$r_{1}(back)$                     & 0.5707($\pm0.0030$)           & 0.6022($\pm0.0057$)     \\
$r_{2}(pole)$                     & 0.2510($\pm0.0115$)           & 0.2585($\pm0.0201$)       \\
$r_{2}(side)$                     & 0.2633($\pm0.0142$)           & 0.2735($\pm0.0257$)      \\
$r_{2}(back)$                     & 0.3114($\pm0.0330$)           & 0.3510($\pm0.1020$)        \\
$f$                               & $41.3\,\%$($\pm$0.9\,\%$$)    & $82\,\%$($\pm$17\,\%$$)  \\
$\Sigma{\omega(O-C)^2}$           & 0.005077                     & 0.004314             \\
\hline
\end{tabular}
\end{center}
\end{table}

\begin{figure}[!h]
\begin{center}
\includegraphics[width=14cm]{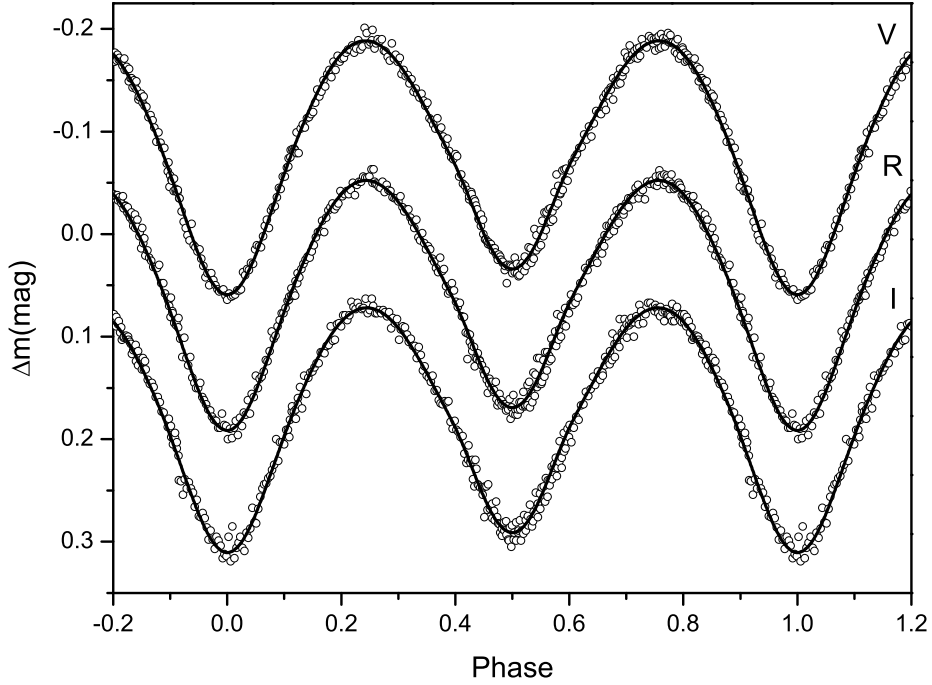}
\caption{Observed (open circles) and theoretical (solid lines) light curves in the $V R_c$ and $I_c$ bands for MQ UMa. The standard deviation of the fitting residuals is 0.006 mag for $V$ band,
 0.007 mag for $R_c$ band and 0.008 mag for $I_c$ band, respectively.}
\end{center}
\end{figure}

\begin{figure}[!h]
\begin{center}
\includegraphics[width=14cm]{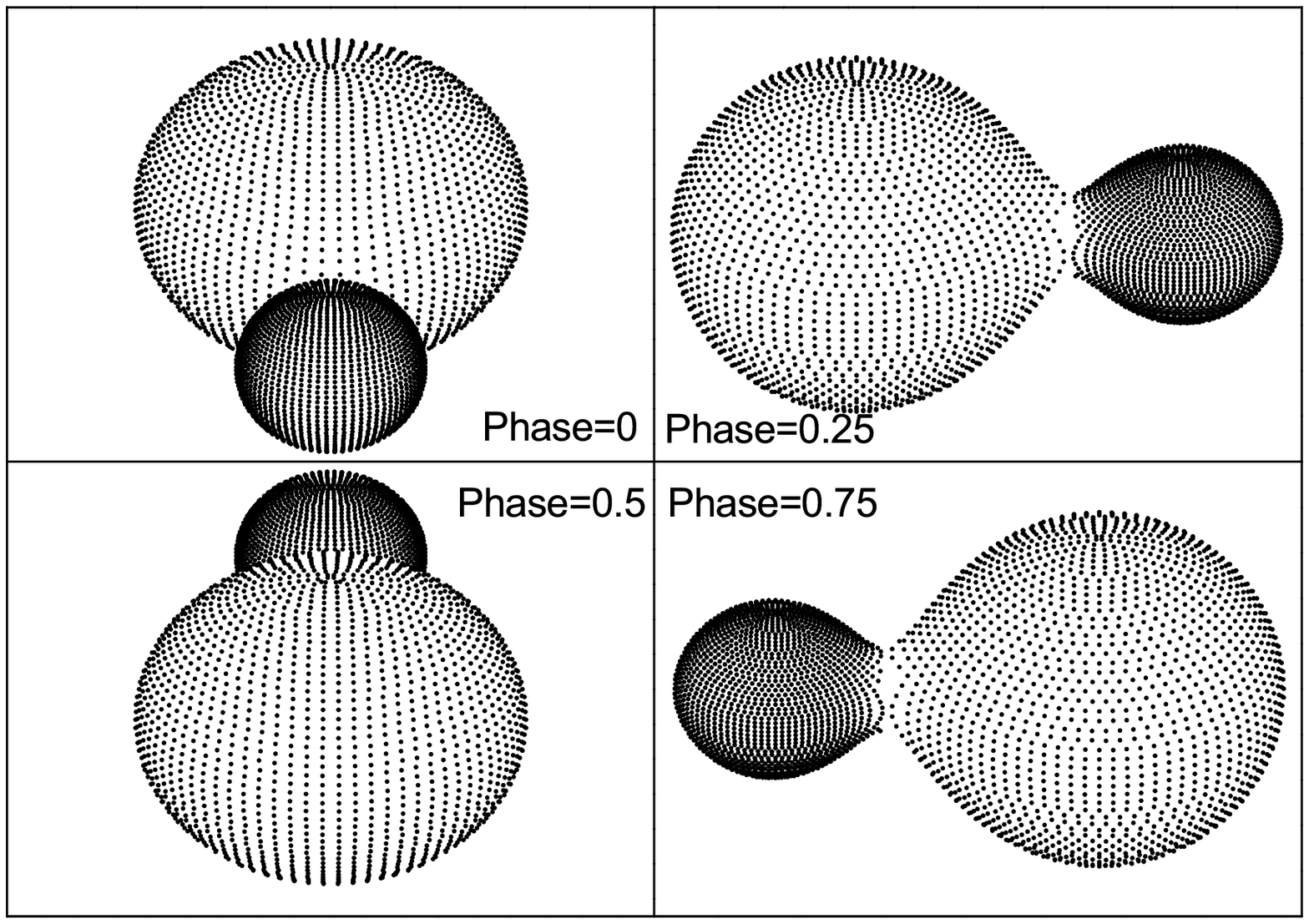}
\caption{Contact configurations of MQ UMa at phase 0.0, 0.25, 0.5, 0.75,}
\end{center}
\end{figure}

To make the photometric solutions of MQ UMa more convinced, we also do $q$-search of MQ UMa by giving the effective temperature of the primary star at $T_1$ = 6100K, $T_1$ = 6300K, $T_1$ = 6500K, $T_1$ = 6700K, respectively. As shown in Fig. 6, although the effective temperature of the primary star ranges from 6100K to 6700K, the shape of q-search curves do not have any significative changes. All of the $q$-search results get the best fitting mass ratio at q = 0.20 $(\pm0.02)$. The convergent photometric solutions (with $l_3$) are listed in Table \ref{rangeT}. As shown in the table, although there is some uncertainty of $T_1$ estimation (6100K to 6700K), the solutions give nearly consistent results, which means the W-D program can tolerate the uncertainty of the primary star's temperature estimated by us. Therefore, the solution of $T_1$ = 6352K is adoptable and we will take the photometric solutions of $T_1$ = 6352K as the final results of MQ UMa hereafter.

\begin{figure}[!h]
\begin{center}
\includegraphics[width=16cm]{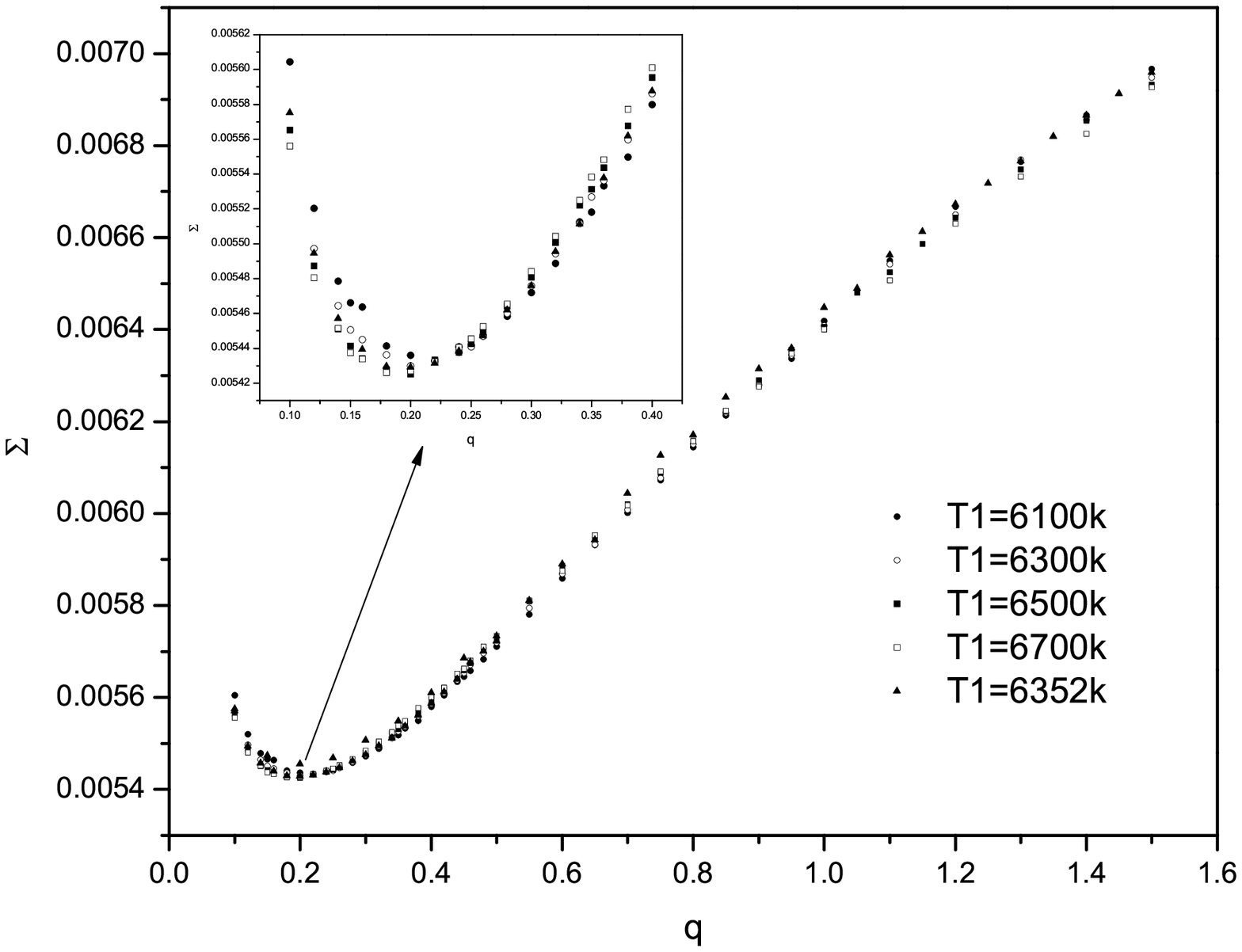}
\caption{q-search diagrams of T1 = 6100K (solid circles), T1 = 6300K (open circles), T1 = 6500K (solid squares), T1 = 6700K (open squares) and T1 =6352K (triangles)}
\end{center}
\end{figure}

\begin{table}[!h]
\caption{Photometric solutions of MQ UMa for different effective temperature}\label{rangeT}
\begin{center}
\scriptsize
\begin{tabular}{lllllllll}
\hline
Parameters                        &    Values                     &     Values                  &       Values                 &      Values                 &    Values       \\
\hline
$T_{1}(K)   $                     & 6700(fixed)                   & 6500(fixed)                 & 6300(fixed)                  & 6100(fixed)                 & 6352(fixed)      \\
q ($M_2/M_1$ )                    & 0.202($\pm0.005$)             & 0.212($\pm0.004$)           & 0.216($\pm0.007$)            & 0.208($\pm0.005$)           & 0.195($\pm0.008$)   \\
$i(^{\circ})$                     & 63.09($\pm0.66$)              & 61.10($\pm0.58$)            & 61.33($\pm0.60$)             & 61.24($\pm0.60$)            & 65.58($\pm0.69$)      \\
$\Omega_{1}=\Omega_{2}$           & 2.158458($\pm0.015663$)       & 2.202068($\pm0.013527$)     & 2.209652($\pm0.018830$)      & 2.193309($\pm0.014472$)     & 2.117244($\pm0.021541$)  \\
$T_{2}(K)$                        & 6496($\pm19$)                 & 6270($\pm18$)               & 6085($\pm16$)                & 5898($\pm16$)               & 6224($\pm25$)        \\
$\Delta T(K)$                     & 204                           & 230                         & 215                          & 202                         & 128                  \\
$T_{2}/T_{1}$                     & 0.970($\pm0.003$)             & 0.965($\pm0.003$)           & 0.966($\pm0.003$)            & 0.967($\pm0.003$)           & 0.980($\pm0.004$)   \\
$\Sigma{\omega(O-C)^2}$           & 0.004314                      & 0.004317                    & 0.004317                     & 0.004321                    & 0.004314             \\
\hline
\end{tabular}
\end{center}
\end{table}

\section{Discussions and Conclusions}
Light curve solutions indicate that MQ UMa is an A-subtype overcontact binary system with a contact degree of $ f=82\,\%$. The two components have nearly the same surface temperature ($\Delta T = 128K$) in spite of their quite different masses and radii, which indicates that the system is under thermal contact. The obtained mass ratio is q = 0.195. We assume that the mass of the primary component is $M_1 = 1.33M_\odot$ \citep{Cox2000} according to its spectral type (F7V), then the mass of the secondary is estimated to be $M_2 = 0.28M_\odot$. The evolutionary status of the primary and the secondary stars are plotted in the H-R diagram as shown in Fig. 7. The evolutionary status of the primary star places it in the middle between the Zero Age Main Sequence (ZAMS) and the Terminal Age Main Sequence (TAMS) lines of the H-R diagram. The secondary component is evidently more evolved than the primary star, and it is clearly overluminous and have higher effective temperature for its present mass.

\begin{figure}[!h]
\begin{center}
\includegraphics[width=14cm]{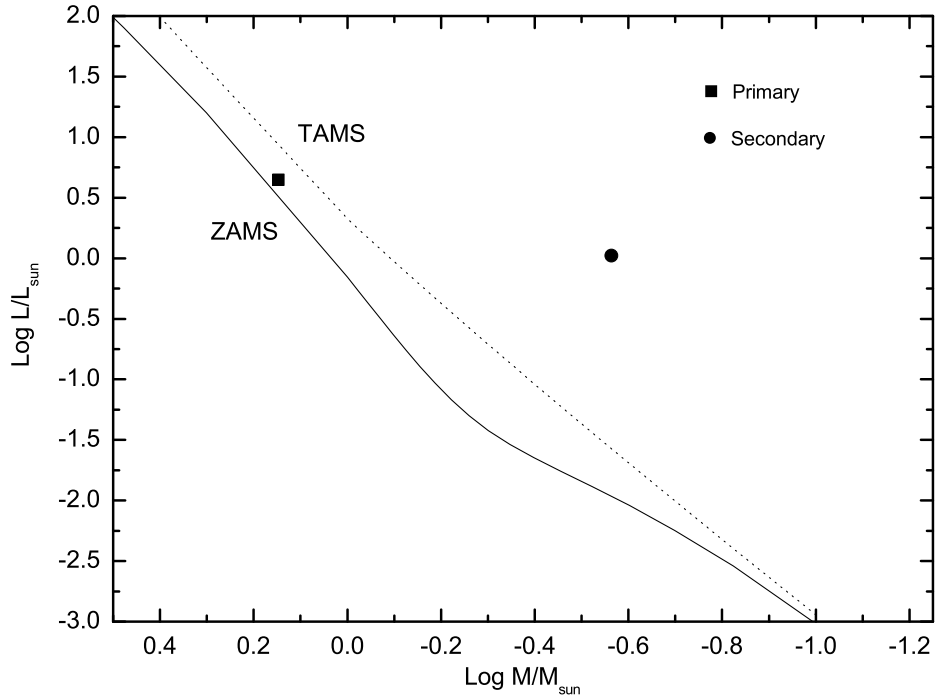}
\caption{The two components of MQ UMa are plotted in the H-R diagram. Solid line represents the Age Main Sequence (ZAMS) and dash line represents the Terminal Age Main Sequence (TAMS).}
\end{center}
\end{figure}

During the photometric processing, the third light ($l_3$) is also included as an adjustable parameter, and the results suggest that the third light contribute nearly a quarter of the total luminosity in the triple system. According to the third light values in $V$ $R_c$ $I_c$ filters listed in Table 4, the color index of the tertiary component are calculated to be $V - R_c = 0.39$ and $R_c - I_c = 0.24$, which corresponds to a spectral type of F5V. It means that MQ UMa has a quite massive and an early type tertiary component. In addition, the existence of the third component may probably be confirmed by spectroscopic observations in the future.  As discussed by \citet{2013ApJS..209...13Q,2014AJ....148...79Q}, the existence of an additional stellar component in the binary system may play an important role for the formation and evolution by removing angular momentum from the central binary system during the early dynamical interaction or late evolution. The angular momentum and orbital period of the binary system will decrease, and the initially detached binaries can evolve into contact configuration via a case A mass transfer during their main sequence evolutionary stage.

High fill-out, low mass ratio overconact binary systems are at the final evolutionary stage of cool short-period binaries. They may merge into a single rapid-rotation star. As a consequence, high fill-out, low mass ratio overconact binary stars may be the progenitors of blue stragglers and FK Com type stars. A sample containing 16 high fill-out, low mass ratio overcontact binaries are presented in Table \ref{overcontact}. Most of them are triple systems and all of them are undergoing a continuous long time period variations (decreasing or increasing). System with a decreasing period will evolve into a single rapid-rotation stars when the photospheric surface of the binary systems is close to the outer critical Roche lobe, while system with an increasing period may merge when it meets the more familiar criterion that the orbital angular momentum is less than 3 times of the total spin angular momentum \citep{1980A&A....92..167H}. As for MQ UMa, the orbital period analysis based on observations collected from public available data and our own observations reveals that it is a triple system. However, we can not determine the exact orbital period of the third component and it is unclear whether it is undergoing a long time period variations (decreasing or increasing). Since the period variations are very important to understand their formation and evolution scenario, those binary systems listed in Table \ref{overcontact} together with MQ UMa will be long-time monitored in the future.

\begin{table}[!h]
\begin{center}
\caption{Parameters of High Fill-Out, Low Mass Ratio Overcontact Binaries}\label{overcontact}
\begin{tabular}{cccccccccccccc}\hline
    Star name   &  $T_1$   & $T_2$    &   Period   &    $q$  &    $f$      &  $i$             & $dp/dt$                 & Cyclic    & Ref.\\
    &           &          &          &    (days)  &         &             &                  &  $\times{10^{-7}}$d/y   &           &      \\\hline
     FG Hya     & $5900K$  & $6012K$  &   0.32783  &  0.112  & $85.6\,\%$  & $85.25^{\circ}$  &     -1.96               &   yes     &   (1)  \\
     GR Vir     & $6300K$  & $6163K$  &   0.34698  &  0.122  & $78.6\,\%$  & $83.36^{\circ}$  &     -4.32               &   yes     &   (2) \\
     IK Per     & $9070K$  & $7470K$  &   0.67603  &  0.191  & $52.0\,\%$  & $77.75^{\circ}$  &     -2.50               &   yes     &   (3) \\
     CU Tau     & $5900K$  & $5938K$  &   0.41254  &  0.177  & $50.1\,\%$  & $73.95^{\circ}$  &     -18.1               &    -      &   (4) \\
     TV Mus     & $5980K$  & $5808K$  &   0.44586  &  0.166  & $74.3\,\%$  & $77.15^{\circ}$  &     -2.16               &   yes     &   (4) \\
     XY LMi     & $6144K$  & $6093K$  &   0.43689  &  0.148  & $74.1\,\%$  & $81.04^{\circ}$  &     -1.67               &    -      &   (5) \\
     V410 Aur   & $6040K$  & $5915K$  &   0.36635  &  0.143  & $52.4\,\%$  & $78.6^{\circ}$   &     +8.22               &    -      &   (6) \\
     XY Boo     & $6324K$  & $6307K$  &   0.37055  &  0.186  & $55.9\,\%$  & $69.0^{\circ}$   &     +6.25               &    -      &   (6) \\
     V857 Her   & $8300K$  & $8513K$  &   0.38223  &  0.065  & $83.8\,\%$  & $85.43^{\circ}$  &     +2.90               &    -      &   (7) \\
     AH Cnc     & $6300K$  & $6265K$  &   0.36044  &  0.168  & $58.5\,\%$  & $90.00^{\circ}$  &     +3.99               &   yes     &   (8) \\
     QX And     & $6500K$  & $6217K$  &   0.41217  &  0.233  & $58.9\,\%$  & $56.20^{\circ}$  &     +2.48               &    -      &   (9) \\
     EM Psc     & $5300K$  & $4987K$  &   0.34396  &  0.149  & $95.3\,\%$  & $88.60^{\circ}$  &     +39.7               &   yes     &   (10) \\
     V345 Gem   & $6115K$  & $6365K$  &   0.27477  &  0.142  & $72.9\,\%$  & $73.3^{\circ}$   &     +0.59               &   yes     &   (11) \\
     V1191 Cyg  & $6500K$  & $6626K$  &   0.31338  &  0.107  & $68.6\,\%$  & $80.4^{\circ}$   &     +4.5                &   yes     &   (12) \\
     CK Boo     & $6380K$  & $6340K$  &   0.35515  &  0.111  & $71.7\,\%$  & $65.9^{\circ}$   &     +0.98               &   yes     &   (13) \\
     DZ Psc     & $6210K$  & $6195K$  &   0.36613  &  0.136  & $89.7\,\%$  & $78.97^{\circ}$  &     +7.42               &   yes     &   (14) \\
\hline
\end{tabular}
\end{center}
\textbf
{\footnotesize Reference:} \footnotesize (1) \citet{2005MNRAS.356..765Q}; (2) \citet{2004AJ....128.2430Q}; (3) \citet{2005AJ....129.2806Z}; (4) \citet{2005AJ....130..224Q}; (5) \citet{2011AJ....141..151Q}; (6) \citet{2005AJ....130.2252Y}; (7) \citet{2005AJ....130.1206Q}; (8) \citet{2006AJ....131.3028Q}; (9) \citet{2007AJ....134.1475Q}; (10) \citet{2008AJ....136.1940Q};
(11) \citet{2009AJ....138..540Y}; (12) \citet{2011AJ....142..124Z}; (13) \citet{2012AJ....143..122Y}; (14) \citet{2013AJ....146...35Y}
\end{table}

\acknowledgments{This work is supported by the Chinese Natural Science Foundation (Grant No. 11133007 and 11325315), the Strategic Priority Research Program ``The Emergence of Cosmological Structure'' of the Chinese Academy of Sciences (Grant No. XDB09010202) and the Science Foundation of Yunnan Province (Grant No. 2012HC011). New CCD photometric observations of MQ UMa were obtained with the 60cm and the 1.0m telescopes at the Yunnan Observatories, and the 85cm telescope in Xinglong Observation base in China. This research has made use of the SIMBAD database, operated at CDS, Strasbourg, France.}

\end{document}